\documentclass[a4,12pt]{article}
\usepackage{bm}
\usepackage{amsmath}
\usepackage{epsfig}
\begin{document}
\begin{center}
{\bf \large Interplay between QCD and nuclear responses.}\\[2ex]

M. Ericson$^{1,2}$, G. Chanfray$^1$\\
$^1$ 
  Universit\'e de Lyon, Univ.  Lyon 1, 
 CNRS/IN2P3,\\ IPN Lyon, F-69622 Villeurbanne Cedex\\
$^2$ Theory division, CERN, CH-12111 Geneva 

\begin{abstract}
We establish the interrelation between the QCD scalar response of the nuclear medium 
and its response to a scalar probe coupled to nucleons, such as the scalar meson 
responsible for the nuclear binding. The relation that we derive 
applies at the nucleonic as well as at the nuclear levels. Non trivial consequences 
follow. One concerns the scalar QCD susceptibility of the nucleon. The other opens 
 the possibility of relating medium effects in the scalar
meson exchange of nuclear physics to QCD lattice studies of the nucleon mass.  
\end{abstract}
Pacs: 24.85.+p 11.30.Rd  12.40.Yx 13.75.Cs 21.30.-x
\end{center}
 \section{Introduction}

The spectrum of scalar-isoscalar excitations is quite different in the vacuum and in the 
nuclear medium.  In the second case it includes low lying nuclear excitations  
and also  two quasi-pion states {\it i.e.} pions 
dressed  by particle-hole excitations. All these lie at lower energies 
than the vacuum scalar excitations which start at $2\,m_\pi$. 
We have shown in previous works \cite{CE03,CEG03,CE05,CDEM06} that this produces a large 
increase of the magnitude of 
the scalar QCD susceptibility over its vacuum value. We have expressed
the origin of this increase as arising from the mixing of the nuclear response to 
a scalar probe coupled to nucleonic scalar density fluctuations into the QCD
scalar response.   

It is natural to investigate also the reciprocal problem of the influence of the QCD 
scalar response to a probe which couples to the quark density fluctuations on the ordinary 
nuclear scalar response of nuclear physics, which is the object of the present 
work. We will study this influence not only for what concerns the  nuclear excitations 
but also for a single nucleon for which only nucleonic excitations are involved. If this 
influence indeed exists, does it lead to non-trivial observable consequences~? We will 
show that this is the case, with two main applications. One concerns the QCD scalar 
susceptibility of a single nucleon. The second is the possibility to infer 
medium effects in the propagation of the scalar meson which binds the nucleus 
from QCD results, such as the lattice ones on the evolution of the nucleon
mass with the pion mass.  
\\

Our article is organized as follows. In  section {\bf 2} we  remind  the mechanisms 
responsible  for the mixing of the nuclear response into the QCD scalar 
susceptibility. We illustrate it in the framework of a nuclear chiral model with a 
scalar and vector meson exchange. We show that this mutual influence also 
exists  at the nucleonic level. In section {\bf 3} we discuss the influence of the quark
structure of the nucleon on the scalar response of nuclear physics in a general framework which 
is able to incorporate also confinement effects.

\section{ Mutual influence of the scalar QCD response and nuclear physics response }
\subsection{Study in a nuclear chiral model}
We first remind how the usual nuclear physics response to a scalar field enters in the 
QCD susceptibility. For this, following ref. \cite{CEG03}, we start from the expression 
of the  modification of the quark  condensate in the nuclear medium, $ \Delta\langle\bar q 
q\rangle(\rho)=\langle\bar{q}q\rangle (\rho)-\langle\bar{q}q\rangle_{vac}$. 
We first use, as in  \cite{CEG03}, its expression for a collection of 
independent nucleons~:
\begin{equation}
 \Delta\langle\bar q q\rangle(\rho)=Q_{S}\,\rho_S.
\end{equation}
where $\rho_S$ is the scalar density of nucleons related to the chemical potential $\mu$ by~:
\begin{equation}
\rho _{S}=4\int \frac{d^3{p}}{(2\pi )^{3}}\, {M\over E_{p}}\, \Theta (\mu -E_{p}) .\label{RHOS}
\end{equation}
We have introduced  the scalar charge of the  nucleon, $Q_S$, proportional to the volume integral of the nucleon 
scalar density of quarks. It is related to the nucleon sigma commutator $\sigma_N$  and the current quark
mass, $m_q$, by~:
\begin{equation}
Q_S ={\sigma_N\over {2m_q}}=\int d^3r \, N|\bar q q({\vec r})- \langle \bar q q\rangle_{vac} | N \rangle .
\end{equation}
The susceptibility of the nuclear medium, $\chi_S^A$, is the derivative of the quark scalar density
with respect to the quark mass. We define it in such a way that it represents a purely nuclear contribution with 
the vacuum susceptibility substracted off ~:
\begin{equation}
\chi_S^A = \left({\partial \Delta\langle\bar q q(\rho) \over \partial m_q}\right)_{\mu}
= \left({\partial (Q_{S}\,\rho_S)\over \partial m_q}\right)_{\mu} . 
\end{equation}
It contains two terms. One arises from the derivative of $Q_S$, which by definition is  the free nucleon QCD scalar 
susceptibility, $\chi_S^N= \partial Q_S/\partial m_q$. The second one involves the 
derivative of the nucleon  density. This term itself is built of two pieces, one involves
antinucleon excitations and is small \cite{CEG03}. The other one, which is larger, involves 
the nuclear response $\Pi_0=-2 M_N p_F/\pi^2$. In this case it is the free Fermi gas one since 
no interactions between nucleons have been introduced. The result  of this derivation is contained 
in the following equation:
\begin{equation}
\chi_S^A = \rho_S\, \chi_N^S \,+ \,2\, Q_S^2\, \Pi_0\,.
\label {CHISGUICHON}
\end{equation}
The nuclear susceptibility is thus the sum of a one-body term and of a  term due the nuclear
excited states, the p-h excitations. This decomposition survives the introduction of the 
interaction, as will be shown next. In this case the free p-h polarization propagator
is replaced by the full RPA one, while the free nucleon susceptibility can undergo
medium modifications and become dependent on the density. \\
The previous result has been generalized in ref. \cite{CE05} to an assembly of nucleons 
interacting through a scalar and a vector meson exchanges, working  at the mean field 
level as in relativistic mean field theories. In this work the condensate was 
obtained as the derivative of 
the grand potential with respect to the quark mass (Feynman-Hellman theorem) and the 
susceptibility as the derivative of the condensate, both being taken at constant chemical potential. 
The result is \cite{CE05}~:
\begin{equation}
\chi  _S=\left( {\partial\langle\bar q q\rangle\over\partial m_q}\right) _\mu
\simeq -2\,{\langle\bar q q\rangle_{vac}^2\over f_\pi^2}\,\left({\partial\bar S
\over\partial c}\right)_\mu .\label{CHIS}
\end{equation}
$\bar S\equiv f_\pi\,+\,\bar s$   is the expectation value of the chiral invariant 
scalar field and $c=f_\pi\,m^2_\pi$ is the symmetry breaking parameter of the model 
used in \cite{CE05}. The quantity $\left({\partial\bar S /\partial c}\right)_\mu$ is related 
to the in-medium sigma propagator~:
\begin{equation}
\left({\partial\bar S \over\partial c}\right)_\mu= -D^*_\sigma= 
{1\over  m^{*2}_\sigma}\,-\,{g^2_S\over  m^{*2}_\sigma}\, \Pi_{S}(0)\,{1\over  m^{*2}_\sigma}
\label{CHISEFF}\end{equation}
where $\Pi_{S}(0)$  is the full scalar polarization propagator, related to the bare one, 
$\Pi_0$  by~:
\begin{equation}
\Pi_{S}(0)={M^*_N\over E^*_F}\,\Pi_0(0)\,
\left[1 -\,\,\left({g^2_\omega\over  m_\omega^2}\, {E^*_F\over M^*_N}\,
- \,{g^{*2}_S\over m^{*2}_\sigma}\, {M^*_N\over E^*_F} \right) \Pi_0(0)\right]^{-1}.\label{PISS}
\end{equation}
In the equations above,  $m^*_\sigma$ is the in-medium sigma mass,  which is obtained 
from the second derivative of the energy density with respect to the order parameter~:
\begin{equation}
m^{*2}_\sigma ={\partial^2 \varepsilon\over\partial\bar s^2}=V''(\bar s)\,+\,
{\partial\left(g_S\,\rho_S\right)\over \partial \bar s}
= m^{2}_\sigma\left(1 \,+\,{3\bar s\over f_\pi}\,+\,{3\over 2}\left({\bar s\over
f_\pi}\right)^2\right)\,
\label{MSIGMA}
\end{equation}
where the potential $V$ responsible for the spontaneous symmetry breaking is the standard 
quartic one of the linear sigma model. In the very last expression of eq. (\ref{MSIGMA})
we have   omitted, as in ref.\cite{CE05}, the small 
antinucleon contribution embedded in the factor $\partial \rho_S/\partial \bar s$. 
Moreover since  for the moment we do not consider, contrary to ref.\cite{CE05}, 
the scalar response of the nucleon due  to confinement, we  also ignore the medium renormalization of $g_S$. The mean scalar field $\bar s$  
being negative, the term linear in $\bar s$  lowers the 
sigma mass by an appreciable amount ($\simeq 30$ \%  at $\rho_0$). This is the chiral 
dropping  associated with chiral restoration   \cite{HKS99} and arising from 
the $3\sigma$ interaction as depicted in fig 1.
  
Since we are interested only in the medium effects the vacuum value of the quantity
$\left({\partial\bar S /\partial c}\right)_\mu=1/{m_\sigma}^2$ has to be subtracted off 
in eq.  (\ref{CHISEFF}) and the purely nuclear suceptibility, $\chi _S^A$,  writes~:
\begin{equation}
\chi _S^A\,=\,{2\,{\langle\bar q q\rangle_{vac}^2\over f_{\pi }^2} }
\left[ {3\,\bar s/ f_{\pi}\,+\,{3\over 2}\left( \bar s/  f_{\pi }\right)^2\,\over  m^{*2}_{\sigma }}
\,+\, {g^2_S\over  m^{*2}_{\sigma}} \,  \Pi_{S}(0)\, {1\over  m^{*2}_{\sigma }}\right] .
\label{CHIAS}
\end{equation}
We see that $ \chi _S^A$ receives two types of contributions, the second being proportionnal
to the full RPA scalar response  $\Pi_{S }$.
The corresponding proportionality factor $r$ between this second contribution and $\Pi_S$ writes, 
to leading order, {\it i.e.}, neglecting the medium modification of the sigma mass~:
\begin{equation}
r = 2\, g^2_S \,{\langle \bar q q\rangle _{vac}^2\over  f_{\pi }^2\, m_{\sigma }^4} \simeq 
2\,( Q^s_S)^2
\label{R}
\end{equation}
where we have introduced the nucleon scalar charge $Q_S^s$ from the scalar field,  
defined below. In the sigma model the free nucleon sigma commutator is the sum  of two contributions, one  
arising from the pion cloud, which depends on the mean value of the squared pion field,
{\it i.e.}, on the scalar number of pions in the nucleonic cloud. In the mean field approximation where 
pion loops are ignored this term does not appear.  The other one, $Q_S^s$, arises from the scalar meson  
\cite{B94,DCE96,SKR06} and is linear in the $\sigma$ field~: 
\begin{equation} 
\label{QSS}
Q_S^s= {\sigma_N^s\over 2m_q} = -{ \langle\bar q q\rangle_{vac}\over f_{\pi }} \int d^3 r\,\langle N| 
\sigma (\vec{r})|N\rangle
 = -{\langle\bar q q\rangle_{vac}\over f_{\pi }}\,{g_S\over m_{\sigma}^2}
\end{equation}
which establishes relation (\ref{R}) if we ignore the in-medium modification of $Q_S^s$,
{\it i.e.}, the difference beween $ m^*_{\sigma}$ and $ m_{\sigma}$.

We now turn to the first part of $\chi_S^A$ which depends on the average scalar field 
$\bar s$. In the low density limit, $\bar s$ reduces to 
$\bar s=-g_S\,\rho_S/ m_{\sigma }^2,$ and we can ignore the term in $\bar s^2$ as well as
the difference beween $ m^*_{\sigma}$ and $ m_{\sigma}$. In this limit the first term
in the expression (\ref{CHIAS}) of  $\chi_S^A$ is linear in the density. 
If we classify it in the decomposition  of eq. (\ref{CHISGUICHON}) for $\chi_S^A$, it obviously  
belongs to the individual nucleon contribution, $\rho_S\,{\chi^N_S},$ to 
the nuclear susceptibility. Writing the linear term  explicitly in eq. (\ref{CHIAS})  we deduce 
the free  nucleon scalar susceptibility from the scalar field, $(\chi^N_S)^s$~: 
\begin{equation}
(\chi^N_S)^s \,=\,-2\,{\langle\bar q q\rangle_{vac}^2\over f_{\pi }^3}\,{3\,g_S
\over m_{\sigma}^4}, 
\label {CHISN}
 \end{equation}
which is   negative ({\it i.e}, it favors an increase in magnitude of the field, similar to paramagnetism). 
It has been obtained here from the low density expression of 
$\chi^A_S$. In fact it can also be obtained directly as the derivative with respect
to the quark mass of $Q^s_S$, the part of the nucleon scalar charge originating 
in the scalar field written in eq. (\ref{QSS}) ~:
\begin{equation}
({\chi^N_S})^s\,= {\partial Q_S^s\over \partial m_{q}} ={\partial \over \partial m_q}
\left( -\,{\langle\bar q q\rangle_{vac}\over f_{\pi}}\,{g_S\over 
m_{\sigma}^2}\right). \label{CHISS2}
\end{equation}
Using the fact that, in the model, $\langle\bar q q\rangle_{vac}/ f_{\pi}$ does not
depend on $m_q$, only the derivative of the sigma mass with respect to $m_{q}$ enters
which, according to the Feynman-Hellmann theorem,  is linked to the sigma commutator, 
$\sigma_{\sigma}$,  of the $\sigma$. In the linear sigma model the derivative with respect 
to the quark mass is replaced by the derivative with  respect to the symmetry breaking parameter, 
$c=f_\pi\,m_\pi^2$, keeping the other original parameters  of the model, $\lambda$ and $v$, constant. 
The result is~: 
\begin{equation}
\sigma_{\sigma }= m_q\,{\partial m_{\sigma}\over \partial m_q} ={3\over 2}\,{m^{2}_{\pi }
\over m_{\sigma}} .
\end{equation}
When inserted in eq. (\ref{CHISS2}), it leads for   $({\chi^N_S})^s$ to    the expression of eq. (\ref{CHISN}).

We now turn to the scattering amplitude for the sigma meson on the nuclear system.
In the same framework  we will first  show that  the  amplitude for the scattering of 
the scalar meson on the nucleon  has the same relation to  the nucleonic susceptibility as 
the case for the nuclear excitation part. Indeed in  the expression 
(\ref{MSIGMA}) of $m^{*2}_{\sigma }$  the  term linear in density is obtained from the 
low density expression~: $3\,\bar s \,m^{*2}_{\sigma }\simeq-(3\,g_S/ f_{\pi} )\,\rho_{S}$.
It represents an optical  potential for the sigma propagation. The corresponding $\sigma N$  
scattering amplitude, $T_{\sigma_N}$, which can also be evaluated 
directly from the graph of  fig. 1, is equal to~:
\begin{equation}
T_{\sigma N}= -3\,g_S / f_{\pi }\label{TSN}.
\end{equation}

\begin{figure}
\begin{center}
\epsfig{file=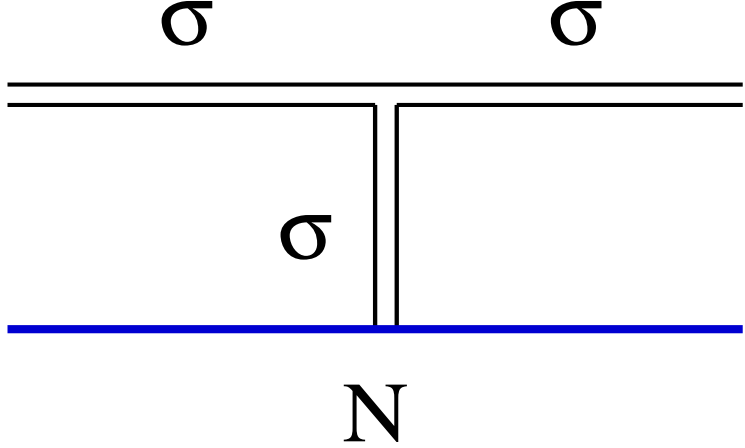,width=10.0cm,height=6.0cm,angle=0}
\end{center}
\caption{Contribution to the sigma-nucleon scattering amplitude responsible for
the lowering of the sigma mass in the medium.}
\label{}      
\end{figure} 

We are now in a situation to relate the nucleon scalar susceptibility (eq. (\ref {CHISN})) 
to the sigma-nucleon amplitude of eq. (\ref{TSN}), with the result~:
\begin{equation}
(\chi_S^N)^s =\,{2\,(Q_S^s)^2\over g_S^2}\, T_{\sigma N}.
\label{CHISNS}
 \end{equation}
The  proportionality factor, $2\,(Q_S^s)^2 / g_S^2$, is the same as for the purely nuclear
excitations. The quantity $g_S$ which appears in this factor factor is due to the $\sigma NN$ 
coupling constant.  Adding the two effects from the nucleonic and nuclear excitations the 
total QCD scalar susceptibility of the nuclear medium (vacuum value substracted) can 
therefore be related to the total response, $T^A$, to the  scalar field through~:
\begin{equation}
\chi _S^A = {2\,(Q_S^s)^2\over g_S^2} T^A\label{RESP}
\end{equation}
where the two members include both the individual nucleon contribution and the
 one arising from the nuclear excitations,with~:
\begin {equation}
T^A= \rho_{S} \,T_{\sigma N}\, +\, g^2_S\, \Pi_{SS} .
\end {equation}
The last term on the r.h.s. represents the (in-medium corrected) Born part of the $\sigma N$ 
amplitude while the first piece represents the non-Born part linked to  nucleonic 
excitations.

Thus there exists a universal scaling factor between the responses of a nuclear or 
nucleonic system to  probes which couple either to the nucleon scalar density
fluctuations or to the quark ones. This relation has allowed us to infer the existence of a 
contribution to the QCD nucleon scalar susceptibility linked to the scalar meson. 
To the best of our knowledge this 
contribution to the nucleon susceptibility has not been discussed previously. It 
 has a link, through the relation (\ref{CHISNS}), to the optical potential 
for the $\sigma$ propagation, which reduces the sigma mass in the medium. 
\subsection{Effect of the two pion propagator}
In order to illustrate the coherence of this approach we will now  extend the previous
description to incorporate the effect of the two-pion propagator, $G,$ which affects the 
nucleon susceptibility in the following way.  The sigma propagator is renormalized by 
the $\sigma$  coupling to  two-pion states,  as discussed in  \cite{CDEM06}.
At zero four-momentum we have~:
\begin{equation}
\label{Dsig0}
- D_\sigma = {1\over m^2_\sigma\,+ \,3\, \lambda \,(m^2_\sigma - m^2_\pi)\, 
{G\over 1\,-\,3\lambda G}}=
{1-3\,\lambda \,G\over \,m^{2}_{\sigma }\,-\,3 \lambda\, m^2_\pi\, G}
\simeq \frac{1}{m^2_{\sigma }} \,-\,{3\,G\over 2\,f^2_\pi} 
\end{equation}
where  $\lambda=(m^{2}_{\sigma }\,-\,m^{2}_{\pi})/2\,f^2_\pi$  and both $D_\sigma$ and $G$ are taken at zero four-momentum. 
In the last term we have restricted to the one pion loop level. We stress that this 
expression only holds for the  sigma, chiral partner of the pion, which is not a chiral 
invariant field. It does not apply to the scalar field responsible for the nuclear 
binding which has to be a scalar invariant (that we have denoted $s$) and which is weakly  
coupled to two-pion states, while the $\sigma$ is strongly coupled. Therefore this treatment  
is done for illustration purpose and not for an application to nuclear physics.

The medium correction to $D_{\sigma}¥$ from the coupling of the $\sigma$ to $2\pi$
states is~: 
\begin{equation}
\Delta D_{\sigma}~=~{3\,\Delta G~ \over 2\,f_{\pi}^{2}},
\end{equation}
where $\Delta G$ is the in-medium modification of the two-pion propagator. In $\Delta G$,  to
lowest order,   one and only one of the two pions of the two-pion propagator  has to be 
dressed by one p-h bubble. It is again possible to interpret the corresponding
modification of the sigma propagator as representing a $\sigma N$ scattering amplitude, 
$T_{\sigma N}^{\pi} $, in which the sigma interacts with the nucleon pion cloud~: 
\begin{equation}
T_{\sigma  N}^{\pi }={3\,m_{\sigma }^4\over 2\,f_{\pi }^2}\,{\Delta G \over \rho _{S}}.
\label{TPISN}
\end{equation}
This is to be compared to the nucleon scalar susceptibility from the pion cloud, which is
\cite{CDEM06}~:
\begin{equation}
\chi _S^N ={3\,\Delta G \over \rho _{S}} \,{\langle \bar{q} q\rangle ^2_{vac}
\over f_{\pi }^4}.\label{CHISPION}
\end{equation}
The proportionality factor between the susceptibility (\ref{CHISPION}) and the 
$\sigma N$ scattering amplitude (\ref{TPISN}) is the same as previously, $2\, Q_S^{s2}/g_S^2$.
We find again that this relation holds not only at the level of p-h excitations
but also for a single nucleon, at the level of the nucleonic excitations which in this 
specific case are of the pionic type.
\\
In summary we have seen that the mixing between the quark density fluctuations  and the 
nucleon ones  implies that the response of a probe which couples to the nucleonic 
density fluctuation  is proportional to the QCD  scalar response. This includes also the 
nucleonic contribution to these responses. As an example we have shown that the chiral 
dropping of the sigma mass in the medium has a counterpart in the form of a contribution 
of the scalar meson to the QCD scalar susceptibility of the nucleon. 
\section{Connection with lattice data}

It is now interesting to connect our results to the available lattice 
simulations  of the evolution of the nucleon mass with the pion mass, equivalently the
quark mass.  At present they do not cover the physical region but only the region  
beyond $m_{\pi } \simeq 400\,MeV$. The derivative 
$ {\partial M_N/ \partial m^2_{\pi}}\, =\sigma_N/m_{\pi}^2$  provides the nucleon sigma 
commutator. In turn the derivative of 
$\sigma_N$ leads to the susceptibility. Both  quantities are  strongly influenced by the 
pion cloud contribution which has a non-analytic behavior in the quark mass, preventing 
a polynomial expansion in this quantity. However the pionic self-energy  contribution to 
the nucleon mass, $\Sigma_{\pi}$, has been separated out in ref. \cite{TGLY04} in a model
 dependent way with 
different  cut-off forms for the pion loops (gaussian, dipole, monopole) with an 
adjustable  parameter $\Lambda$. The remaining part  is expanded in terms of $m^2_{\pi}$  
as follows:
\begin {equation}
M_N(m^{2}_{\pi}) = 
a_{0}\,+\,a_{2}\,m^{2}_{\pi}\, +a_{4}\,m^{4}_{\pi}\,+\,\Sigma_{\pi}(m_{\pi}).
\label{EXPANSION}
\end{equation}
The best fit value of the parameter $a_{4}$  which fixes the  susceptibility shows 
little sensitivity to the shape of the form  factor, with a value 
$a_4 \simeq-\,0.5\, GeV^{-3}$ while $a_2 \simeq 1.5\,GeV^{-1}$ (in  a previous work 
\cite{TLY04} smaller values of $a_2$ and $a_4$ were given~: $a_2\simeq 1\,GeV^{-1}$ and 
$a_4 \simeq-\,0.23\, GeV^{-3}$).  We can infer 
the non-pionic pieces of the sigma commutator and of the susceptibility from the 
expansion (\ref{EXPANSION})~: 
\begin{equation}
\sigma_N^{non-pion} = m^2_{\pi} \,{\partial M\over \partial m^2_{\pi }}
~=~a_2\, m^2_{\pi}\, +\, 2\,a_4\,  m^4_{\pi}~\simeq 29\,MeV \,.
\label{SIGMALATTICE}
\end{equation}
It is largely dominated by the $a_2$ term.
The corresponding value for $a_2\simeq 1\,GeV^{-1}$ is $\sigma_N^{non-pion}=20\,MeV$.

In turn the nucleon susceptibility is~:
\begin{equation}
\chi_S^{ N, non- pion}~= 2\,{\langle\bar q q\rangle_{vac}^2 \over f^4_{\pi} }
{\partial \over \partial  m^2_{\pi }}\left({\sigma_N^{non-pion} \over m^2_{\pi }}\right)~=  
{\langle\bar q q\rangle_{vac}^2 \over f^4_{\pi} }\,4\,a_{4}
 ~\simeq -5.4\, GeV^{-1} 
\label{CHILATTICE}
\end{equation}
The non-pionic susceptibility is found with a negative sign, as expected from the scalar 
meson term. In ref. \cite{TGLY04} however, the negative sign is interpreted differently. It is attributed 
to possible deviations from the  Gellman-Oakes-Renner (GOR) relation which links quark and pion 
masses. 

It is then interesting to test if the empirical values from the lattice are compatible 
with a pure scalar meson contribution. We thus tentatively make the following 
identifications~:
\begin{equation}
Q_S^s~= {\langle\bar q q\rangle_{vac}\over f_{\pi }}\,{g_S\over m_{\sigma}^2}~=
 {\sigma_N^{non-pion}\over (2\, m_q)}\simeq 2.4,
\end{equation}
with $2\,m_q=12 \,MeV$ (taking 
$a_2\simeq 1\,GeV^{-1}$ one would get $Q_S^s=1.66$). It is interesting to translate 
this number into the value of the mean scalar field in the nuclear medium which, to leading 
order in density, is~:
\begin{equation}
-\bar s= {g_s\,\rho_S\over m^2_\sigma}={Q_S^s\,f_{\pi }\,\rho_S\over\langle\bar q q\rangle_{vac}}
={\sigma_N^{non-pion}\over (2\, m_q)}\,{f_{\pi }\,\rho_S\over\langle\bar q q\rangle_{vac}}
={a_2\,+\,a_4\,m_\pi^2\over f_\pi}\,\rho_S~.
\end{equation}
 At normal density the value $|\bar s(\rho_0)|\simeq 21\, MeV$, quite 
compatible with nuclear phenomenology. The second identification concerns the 
susceptibility. If the non pionic susceptibility would arise entirely from the scalar 
field, we should have~: 
\begin{equation}
 \chi_S^{N, non- pion}~=~-{2 \,(Q_S^s)^2\over g_S^2}\,{3\,g_S\over f_{\pi}} 
 =~{\langle\bar q q\rangle_{vac}^2 \over f^4_{\pi} }\,4\,a_{4} 
\end{equation}
which would give, using the GOR relation~:
\begin{equation}
-a_4={3\over2} \,{(\sigma_N^{non-pion})^2 \over g_S\,f_\pi\,{m_\pi}^4} = 3.1\, GeV^{-3},
\end{equation}
much larger than the lattice value, $-a_4= 0.5 \, {\rm GeV}^{-3}$. Again,  taking  
$a_2\simeq 1\,GeV^{-1}$ one would get $-a_4\simeq 1.2\, GeV^{-3}$, still larger in magnitude than the 
corresponding lattice value $(-0.23 \, GeV^{-3})$.  Thus the linear $\sigma $ model 
which fails  to account for the saturation properties, due to the excess attraction 
produced by the chiral softening of the sigma mass,  also leads to too large a 
susceptibility from the scalar meson. In fact in  this work we show  that the two problems 
are linked. Some mechanism  suppresses the chiral softening of the 
$\sigma$  mass as well as the large nucleonic susceptibility from the scalar meson, 
incompatible with lattice data. It is indeed likely that the scalar meson is not the 
only non-pionic contribution. In  a previous work \cite{CE05} we have invoked confinement 
and the quark meson  coupling model (QMC) \cite{G88,GSRT96} as a source of cancellation 
for the chiral softening of the sigma mass. It turns out that it has also a cancelling 
effect in the nucleon scalar susceptibility. Indeed,  for 3 valence quarks confined in a bag of radius R,  Guichon \cite{G05} derived 
$\chi_S^{N, bag }\simeq\,+\,0.25\, R \simeq 1\,GeV^{-1}$, for a value  $R=0.8\,fm$.  
Translated into the parameter $a_4$, one has  $a_4^{bag}\simeq \, + \,0.1\, GeV^{-3}$.
Contrary  to the other components which are negative (of paramagnetic nature), it has 
a positive sign (of 
the diamagnetic type, linked to quark-antiquark excitations). The bag susceptibility 
indeed produces a mild cancellation effect.  

It is then natural to try to extend the linear sigma model description so as to 
incorporate 
other effects than the scalar meson ones (or chiral symmetry breaking effects), 
for instance those arising from confined valence quarks.  

 \section{ Generalization and implications for nuclear physics}
\subsection{ General relation }

In view of the limitations of the linear sigma model discussed previously,
 a more general approach is desirable. 
The aim is to link the response of the nuclear medium to the  scalar 
nuclear field and the QCD responses, in such a way that both quantitites include all the 
components of the individual nucleon contribution whatever their origin. 
Of course it is not possible to achieve this goal without some assumptions on the nature of 
the probe. We keep the basic assumption that the scalar field which couples to the
nucleons, couples
to the quarks of the nucleon condensate, as is the case in the linear sigma model.
Thus its presence can induce a readjustement of the quark structure of the nucleon, that we
evaluate in the way described  below.

Consider a nuclear medium with a scalar nucleon density $\rho_S$.  
By definition the response of this medium to a scalar field which couples to the nucleon 
scalar density fluctuations (with a unit coupling constant)  is  the change in
the nucleon scalar density for a small change of the nucleon mass. It is 
$\Pi_S =({\partial \rho_{S}/ \partial M_N})_\mu$, the derivative being taken at 
constant chemical
potential. With a coupling constant $g_S$ this result should be multiplied by $g_S^2$. In 
the free Fermi gas case this derivative leads the the quantity $-2\,Mp_F/\pi^2$,
the free Fermi gas response.  For nucleons interacting via $\sigma $ and $\omega $ 
exchange, the expression of the scalar nucleon density is~:
\begin{equation}
\rho_S=\int\,{4\,d^3 p\over (2\pi)^3}\, {M^*_N\over E^*_p}\,
\Theta\left(\mu\,-\,E^*_p\,-\,{g^2_{\omega }\over m_{\omega }^2}\,\rho \right).
\end{equation}
where $ M^*_N~=~M_N~(1+\bar{s}/{f_\pi})$ is the nucleon effective mass, linked to the mean
scalar field $\bar{s}$ and 
$ E^*_p~=~\sqrt{p^2~+~M^{*2}_N}$. 
The mean field $\bar{s}$  is obtained from the minimization equation of the energy density
 $\epsilon$ ~
:\begin{equation}
{\partial\varepsilon\over \partial\bar s}= g_S\,\rho_S\,+\,V'(\bar s)=0\, .
\label{MIN}\end{equation}
It is then possible to check that the derivative of the scalar nucleon density with
respect to the nucleon mass leads to the full RPA scalar polarization propagator,
$\Pi_{S}$, as  defined in eq. (\ref{PISS}).
In this expression of the response as the derivative of the nucleon density the nucleon 
structure is not incorporated. It only includes the effect of the nuclear excitations and 
not that of the nucleonic ones. In order to include them we have to account for the 
internal nucleon structure, {\it i.e}., the quark structure. It is the quark medium, and not 
only the nucleon one, 
which  responds to the same excitation, {\it i.e.}, to the modification of the nucleon  
mass   $\delta M_N$.  Accordingly we make the following conjecture, writing 
the full response $\mathcal{R_S^A}$  as~:
\begin {equation}
\mathcal{R_S^A}=~ {1\over 2\,Q_S}\,\left({\partial  \rho_S^q\over \partial M_N}\right)_\mu
\end {equation}
where  $\rho_S^q$ is the quark scalar density and the factor ${1/ 2\,Q_S}$ in front of 
the derivative is put for normalization  purpose. Each nucleon containing a scalar number 
of quarks $2\,Q_S= \sigma_N/m_q$, the scalar density of quarks is  
$\rho_S^q = 2\,Q_S \,\rho_S $. The derivative  involves two terms~:
\begin {equation}
\mathcal{R_S^A}=~
 {1\over Q_S}\,\left({\partial \over \partial M_N}(Q_S\,\rho_S)\right)_\mu=~ 
\left({\partial\rho_{S}\over \partial M_N}\right)_\mu \,+\,
 {\rho_{S}\over2\, Q_S^2}\, {\partial Q_S\over \partial m_q} .
\end {equation}
In the last term we have replaced the derivative with respect to the nucleon mass by the 
one with respect to the quark mass, with ${\partial M_N/ \partial m_q} =2\,Q_S$,  
which introduces the nucleon susceptibility $\chi_S^N$. The overall result writes~:
\begin {equation}
\mathcal{R_S^A}~=~ {\chi_S^N\over2\, Q_S^2} \, \rho_{S}\,+\,\Pi _S .
\label{RSA}
 \end {equation} 
The interpretation of this equation is clear. This decomposition is obvious and analogous 
to the one of eq. (\ref{CHISGUICHON}). The term linear indensity represents the individual 
nucleon response from the nucleonic excitations, while the term in $\Pi_S$
embodies nuclear excitations. The new information is that the single nucleon response is 
proportional to the QCD one, $\chi_S^N$, with the same proportionality factor $1/(2Q_S^2),$ 
as was found previously for the nuclear excitations. All in all, the eq. (\ref{RSA})
writes~: 
\begin {equation}
\mathcal{R_S^A}~=~ {1\over2 \,Q_S^2}\,\chi_S^A\label{RAS}
\end{equation}
where  $\chi_S^A$ represents the total scalar QCD susceptibility of the nuclear medium 
(vacuum value substracted) and both members incorporate the individual nucleon 
contribution. This results holds for a unit coupling constant. For a coupling constant 
$g_S$  (as is the case for the nuclear scalar field)  the r.h.s should be multiplied by 
$g_S^2$. Accordingly the corresponding $\sigma N$ amplitude is ~:
\begin{equation}
T_{\sigma N} =  {\chi_S^N g_S^2\over2\, Q_S^2} .
 \label{TTOTAL}
\end{equation}

We will now comment this result and we then will apply it to the problem of the propagation 
of  the scalar field which mediates the nuclear attraction. 
Our relation (\ref{RAS}) has a close resemblance to the 
previous one, (\ref{RESP}), derived in the linear sigma model but here we do not inquire
about the origin of the terms, $\chi_S^N$ and $ Q_S$. With the values of the linear
 sigma model for these quantities we recover the previous result of this model. 

Our relation  (\ref {TTOTAL}) is also very similar to the one of  the quark-meson coupling model \cite{G88}. In QMC, the bag positive 
susceptibility manifests itself in the form of a repulsive interaction in the 
propagation of the scalar nuclear field. The corresponnding scattering amplitude  is 
related to the bag suceptibility by a relation identical to our eq. (\ref{TTOTAL}), but
  only bag 
quantities appear and the scalar charge entering this relation is that of the bag, 
which is $Q_S^{bag}\simeq 0.7$. As QMC does not incorporate the  chiral potential which
implies  the three-scalar coupling, only the repulsive three-body interaction from the 
bag  structure enters. Our expression  (\ref {TTOTAL}) thus covers the two extreme situations, when the nucleon 
mass originates totally from the condensate as is the case in the $\sigma$ model, or when it is only due to confinement.
It is legitimate to believe that it is able to describe a more general situation with a
mixed origin. 

\subsection{ Illustration in a hybrid model of the nucleon}
In the following we will illustrate the relation  (\ref {TTOTAL}) in a model of the nucleon proposed by 
Shen and Toki \cite{ST99}, where the nucleon mass originates in part from its 
coupling to the condensate and in part from confinement. It consists in the
following: three constituant quarks, described in the Nambu-Jona-Lasinio model (NJL), are kept together by a
central harmonic force so as to mimick confinement.
We have chosen for simplicity the form: $((K/4)(1+\gamma_0)\,r^2$ which leads to analytical results.
Although oversimplified the model  gives an intuitive picture  of the role played by confinement. Denoting M the mass of a free constituant
quark and E that of the bound one, the nucleon mass is given~:
 \begin{equation}
M_N =3\,E=3\left( M+{3\over2}\sqrt{K\over E+M} \right) \,.
\end{equation}
It is increased as compared to the value, $3M$, for three independent constituant quarks.. 
The nucleon scalar charge, $Q_S$, is~:  
\begin{equation}
 Q_S ={3\over 2}\, {\partial E\over \partial m_q}  
={3\over 2} \,{\partial E\over \partial M}  {\partial M\over \partial m_q}
\end{equation}
with :
\begin{equation}
 {\partial E\over \partial M}=c_S
={E+3M\over3E+M}\,.
\end{equation}
As $E>M$, $c_S<1$, the nucleon scalar charge is reduced as compared to a collection of three independent constituant quarks.
The nucleon scalar susceptibility, $\chi^N_{S}$, given by the next derivative, is composed of two terms arising respectively 
from the derivative of $c_S$ and from that of  $\partial M/ \partial m$. The second
part leads to the susceptibility, $\chi^q_{S}$, of a free constituant quark,
\begin{eqnarray}
\chi^N_{S} = 
 {\partial Q_S\over \partial m_q}= {3\over 2} 
\left[ {\partial c_S\over \partial M}\left({\partial M\over \partial m_q}\right)^2
+ c_S \, {\partial^2  M\over \partial^2  m_q^2} \right] \\ \nonumber
 = {3\over 2}\, {\partial c_S\over \partial M}\left({\partial M\over \partial m_q}\right)^2
+ 3\,c_S \,\chi^q_{S}
\end{eqnarray}
with:
\begin{equation}
 {\partial c_S\over \partial M}={24 \,(E^2- M^2)\over {(3E+M)}^3}\,.
\end{equation}
Notice that this last derivative is positive since $E>M$ and that it vanishes in the absence of confining force, when $E=M$. 
Therefore the first part of the expression of $\chi^N_{S}$ represents the part of the susceptibility originating in confinement and, as in QMC, 
it is positive.

The scalar coupling constant $g_S$ is linked to the derivative of the nucleon mass with respect to the mean scalar field $\bar s$:
\begin{equation} 
g_S = 3\,{\partial E\over \partial \bar s}= 3\,  {\partial E\over \partial M}{{\partial M\over \partial \bar s}}
= 3 \,c_S\, g_{q }
\end{equation}
where $g_{q}$ is the corresponding coupling constant for a constituant quark. 

The nucleon response to the scalar field originating in confinement, $\kappa _{NS}$, is linked to the second derivative of the nucleon mass with respect to the scalar field~: 
\begin{equation} 
\kappa _{NS} = 3\,
{\partial^2 E\over \partial \bar s^2}= 3\, 
 {\partial  c_S\over \partial M}\left(
{\partial M\over \partial \bar s}\right)^2\,.
\end{equation}
The ratio between the part of the nucleon scalar susceptibility which is due to confinement and $\kappa_{NS}$ is
\begin{equation}
r={1\over 2}\,{({\partial M\over \partial m_q})^2\over ({\partial M\over \partial s})^2}
= 2 \,{Q_S^2 \over g_S^2},
\end{equation}
the same ratio as was previously found.
   
As for the scalar $\sigma N$ amplitude from the tadpole term, $T_{\sigma N} =3\,g_S/f_{\pi}$,
 it should be compared to the other part of the susceptibility. We define $r'$ as the
 corresponding ratio through~:
\begin{equation}
{3\over 2}\, c_S\, {\partial^2  M\over \partial m_q^2} = r'\,{3\,g_S\over f_{\pi}}\,. 
\end{equation}
In the semi-bosonized version of the NJL model we have~:
\begin{equation}
{\partial  M\over \partial  m_q}  =- {2 \,{g_q\,\langle\bar q q\rangle_{vac}
\over f_{\pi}\,m^2_{\sigma}}} 
\end{equation}
and  
\begin{equation}
{\partial^2  M\over \partial  m_q^2} = 2 \,\chi^q_{S} = {2 \,g_q\,\langle\bar q q\rangle_{vac}^2 \over 
{f_{\pi}^3}\,m^4_{\sigma}}
\end{equation}
 in such a way that the ratio $r'$ becomes~: 
\begin{equation}
r'= {2\, \langle\bar q q\rangle_{vac}^2 \over 
{f_{\pi}^2}\,m^4_{\sigma}} = {2\,Q^2_S\over g^2_S} { \equiv r}.
\end{equation}
Since the same ratio applies to the two parts, $r'\equiv r$, it can be factorized so as
to obtaIn the relation (\ref{TTOTAL}), which is thus confirmed in this hybrid model.
\\
Numerically a value of the ratio $E/M \simeq 2.1$, which gives $c_S\simeq 0.7$, leads to a reasonable value for 
$g_A$. It results in a value of the dimensionless parameter $C=(f^2_{\pi}/ 2M)\,\kappa_{NS} \simeq 0.1$, 
while the value needed to account for the saturation properties  in the framework of chiral models is $C\simeq 1$ \cite{CE07}. 
Even if it fails to account for the numerical value this model has the merit to confirm the validity of the relation 
between the QCD response and the one to the nuclear scalar field in a situation where confinement enters
and to illustrate the role played by confinement, with the introduction of a positive component in the susceptibility 
which opposes an increase of the nuclear scalar field.

We can now turn to the quantitative applications of the relation  (\ref{TTOTAL}). Since both the total (non-pionic) 
nucleonic susceptibility and  scalar  charge enter in the expression of the (chiral 
invariant)  scalar-nucleon scattering amplitude, it is legitimate to use for these two 
quantities the phenomenological values obtained from the  lattice data. We can 
therefore infer the medium effects in the  propagation of the s field from the lattice 
results of eq. (\ref{SIGMALATTICE}) and (\ref{CHILATTICE}) as~:
\begin{equation}
-D_s^{-1}= 
m_\sigma ^2\,+\,{g_S^2\over2\, Q_S^{2}}\,\chi^N_{S}\,\rho_{S}
=m_{\sigma }^2\,+\,g_S^2\,{2\, a_4\over {(a_2\,+\,2\,a_4\,m_{\pi }^2)}^2}\, \rho _S
\end{equation}
where in the first equation only non-pionic quantities enter, hence the introduction
of the parameters  $a_2 $ and $a_4$ which have been defined in the lattice expansion.
Numerically, at normal nuclear density, and for a value of the coupling constant 
$g_S=10 $, the second term on the rhs of the second equation takes the value $0.06 \,GeV^2$ 
 (a similar value is found for the other set of parameters $a_2$ and $a_4$).
For a sigma mass of $m_\sigma=0.75\, GeV$,  this represents at $\rho_0$ only a 6\% decrease of the mass,
much less that the chiral dropping alone and  in much better agreement with 
the nuclear phenomenology \cite{CE05,CE07}. 
\section{Conclusion}
In summary we have studied in this work the interplay between the nuclear responses to  
probes which couple either to nucleon or to quark scalar density fluctuations. We have
found that the two responses are closely related, being proportional to each other. The 
scaling coefficient involves the scalar charge of the nucleon $Q_S$. Our result holds not
only at  the level of the nuclear excitations but also at the 
nucleonic ones such that both responses incorporate the individual nucleon contributions to the 
nuclear response. Thus the scalar response of a nucleon to the nuclear scalar field  is 
proportional to its QCD scalar susceptibility. We have confirmed this relation in the Shen and 
Toki model of the nucleon where its mass arises in part from the coupling to the condensate and in 
part from confinement.

One application of this relation concerns a free nucleon. The $\sigma N$ amplitude from the tadpole term  
has  a counterpart in the QCD scalar susceptibility in the form of a negative 
contribution beyond the pionic one. We have tested its existence in the lattice results on the nucleon mass 
evolution with the pion mass, as analyzed in ref. \cite{TGLY04}. The expansion of ref.  \cite{TGLY04}  is indeed 
compatible with a negative component for the non-pionic susceptibility. However the magnitude does not fit,
indicating the existence of other components, with a cancelling effect. In fact a similar 
cancellation has to occur in the saturation problem of nuclear matter. The  $3\sigma$
coupling is responsible for a lowering of the $\sigma$ mass, which produces too much
attraction at large densities and destroys  saturation. It has to be compensated. In the optics
of the present work, the two effects are related. The full $\sigma N$ scattering amplitude
being proportional to the susceptibility, a cancelling effect in the amplitude is
automatically reflected in the susceptibility. Confinement may be invoked as a natural
mechanism for cancellation.

The existence of a link between QCD and nuclear physics quantities allows the
derivation of parameters of the $\sigma \omega$ model from the lattice results
on the nucleon mass dependence on the quark mass. This procedure leads to a mean scalar
field $|\bar s (\rho_0)|\simeq 20\,MeV$. In our approach the nucleon response to this field can also 
be derived from the lattice data. Altogether this method leads to a satisfactory description of the nuclear matter 
saturation properties.

\end{document}